\documentclass[a4paper,12pt]{article}
\usepackage{amsmath}
\usepackage{amssymb}
\usepackage{latexsym}
\newcommand{\C}{\mathcal}
\newcommand{\nn}{\nonumber}
\newcommand{\G}{\Gamma}

\renewcommand{\d}{\partial}
\renewcommand{\vec}[1]{{\bf#1}}

\begin{document}
\title{
Constraining a possible dependence of Newton's constant on the Earth's
magnetic field}
\author{Andreas Rathke\\
ESA Advanced Concepts Team (SER-A)\\ ESTEC, Keplerlaan 1,
2201 AZ Noordwijk, The Netherlands\thanks{e-mail: andreas.rathke@esa.int}
}

\maketitle

\begin{abstract}
  Some time ago Mbelek and Lachi\`eze-Rey proposed that the
  discrepancy between the results of the various measurements of
  Newton's constant could be explained by introducing a gravielectric
  coupling between the Earth's gravitational and magnetic fields
  mediated by two scalar fields. A critical assessment of this model
  is performed. By calculating the static field configuration of the
  relevant scalar around a nucleus in the linearised theory and then
  folding this result with the mass density of the nucleus its
  effective gravitational mass is determined. Considering test bodies
  of different materials one finds violations of the weak equivalence
  principle for torsion-balance experiments which are four orders of
  magnitude beyond the current experimental limit, thus rendering the
  model non-viable. The method presented can be applied to generic
  theories with gravielectric coupling and seems to rule out in
  general the explanation of the discrepant measurements of Newton's
  constant by such couplings.

\end{abstract}


\section{Introduction}

The values of Newton's constant, obtained by various experiments, differ by
the order of $10^{-3}$ (cf.\ \cite{Gillies} for an overview). This
discrepancy is understandable from the field-theoretical point of view
considering the extraordinary weakness of the gravitational
interaction. Nevertheless the
experimental situation is unsatisfactory because the discordance of
measurements lies beyond the experimental errors given by the various groups.
Hence it seems a legitimate hypothesis that the mismatch between the
experimental values might be an indiction of ``new physics''. A proposal
towards this direction has been put forward by Mbelek and Lachi\`eze-Rey in
\cite{ML} and has recently been further elaborated by them to include as a
cosmological implication the variation of the fine-structure constant
\cite{MLfine}.

The model of \cite{ML} is based on an action which is claimed by the
authors to consists of the
action of dimensionally reduced five-dimensional Kaluza-Klein (KK)
theory amended by 
matter fields and a second four-dimensional scalar field. The KK action
provides the four-dimensional gravitational field, the electromagnetic field
and a massless KK scalar. In contrast to conventional KK theory both the KK
scalar as well as the second scalar are coupled to the electromagnetic field
and the other matter fields in a way which cannot be reduced to a
universal coupling.

The non-universal coupling between the electromagnetic field and the
KK scalar leads to a dependence of the effective gravitational
constant on the electromagnetic field. Mbelek and Lachi\`eze-Rey
investigated the effect of the Earth's magnetic field on the
gravitational constant in view of this gravielectric coupling and,
employing a dipole fit of the Earth's magnetic field, concluded that
this new coupling could not only explain the discrepancy between the
various measurements of Newton's constant but is also statistically
preferred compared to the assumption of a constant gravitational
coupling.  In conclusion, the magnetic field of the Earth would lead
to a modification of the gravitational coupling of the order of
$10^{-3}$.  Conflict with the lack of gravitational anomalies in
astrophysical sources, e.\,g.\ due to the Sun's magnetic field or in
neutron stars, are avoided by assuming a temperature dependence of the
gravielectric coupling leading to its vanishing at high
temperatures. On the other hand, taking into account the extraordinary
weakness of the Earth's magnetic field compared to the magnetic fields
producable in laboratories, the result of \cite{ML} is not only of
interest from the fundamental physics point of view but also seems to
open the door to a whole new technology of gravity
engineering. Considering, furthermore, the echo the model has found in
popular science publications (see e.\,g.\ \cite{press}) a thorough
investigation of further implications of the model appears highly
desirable.

Unfortunately, a closer analysis shows that the model is already ruled
out by current experimental constraints. Due to the complicated
coupling of both scalars (and the metric) to the electromagnetic field
the model falls into the class of non-metric theories and will
inevitably exhibit violations of the weak equivalence principle
(WEP).\footnote{cf.  \cite{Will} p.\ 22 ff.\ for the definition of the
term ``non-metric theory''.} Since the model under consideration is a
biscalar-tensor theory there are, however, no well established phenomenological
constraints on its coupling parameters, like in the case of ordinary
scalar-tensor theories.\footnote{cf.\ \cite{Willlife} for the constraints
on scalar-tensor theories.} Hence we resort to demonstrating the
non-viability of the model for the special situation of an
E\"otv\"os experiment.

As could have been anticipated from the
surprisingly strong gravielectric coupling, one finds a strong
modification of the effective gravitational constant in the
electric field of nucleus. The resulting modification of the ratio
between the inertial and the effective gravitational mass is
proportional to $(Z/A^{1/3})^2$ where $Z$ is the proton number of the
nucleus and $A$ is its nucleon number. The resulting violations of the
WEP are approximately four orders of magnitude beyond current
experimental bounds from E\"otv\"os experiments. Consequently, the model of
\cite{ML} is not viable. 

The layout of this treatise is the following: In the next section we discuss
the underlying action of the model of \cite{ML}. We find that the action of
model \cite{ML} is \emph{not} as claimed the effective action of
dimensionally reduced KK theory plus additional matter fields. The KK part of
the action has an additional kinetic term for the scalar which is not
obtainable by dimensional reduction of five-dimensional KK theory. Also the
relative signs of the various terms in the action turn out not to be
consistent.  In particular the action of \cite{ML} has a negative
stress-energy tensor for the electromagnetic field. We thus correct this
inconsistency of the model by carrying out the necessary sign changes and take
the resulting action as a starting point for our considerations. In Sec.\ 
\ref{lin} we derive the weak-field approximation of the model by linearising
its action and deriving the linearised equations of motion.  Starting from the
static limit of the equations of motion we obtain the solution for the
linearised KK scalar field in a nucleus in Sec.\ \ref{WEP} by a semiclassical
treatment of the nucleus.  Inserting this solution in the static metric
perturbation of a nucleus we determine the strength of violation of the WEP in
the experiments \cite{Braginsky,Roll,Su}. In Sec.\ \ref{concl} we summarise
our results and discuss the place of the model in the classification of test
theories of gravity. We also comment on the generic incompatibility between a
strong gravielectric coupling and the maintenance of WEP constraints.
  The appendix further discusses the inconsistency in the
original action and spots part of its origin in the paper \cite{MLghost}.

\section{Inconsistency of the underlying action}

Mbelek and Lachi\`eze-Rey in \cite{ML} discuss properties of the action
\begin{equation}
S_\text{ML} = S_\text{``KK''} + S_\Psi + S_{\rm matter} \, .
\label{ac1}
\end{equation}
In Eq.\ (\ref{ac1}) $S_\text{``KK''}$ (the meaning of the quotation marks will
become clear below) denotes the action
\begin{equation}
S_\text{``KK''} = - \int d^4x\, \sqrt{-g} \left[ 
\frac{c^4}{16\pi G_0} \Phi R + \frac 14 \epsilon_0 \Phi^3 F_{\mu\nu} F^{\mu\nu}
+ \frac{c^4}{4\pi G_0} \frac{\d_\mu \Phi \, \d^\mu \Phi}\Phi
\right] \, ,
\label{MLKKaction}
\end{equation}
where $G_0$ denotes the effective four-dimensional gravitational
constant in the Einstein frame. Eq.\ (\ref{MLKKaction}) is claimed to be
the reduced effective four-dimen\-sional action of five-dimensional
KK theory in the Jordan frame.  The part of the action,
Eq.\ (\ref{ac1}), denoted $S_\Psi$ is the action of a scalar field
$\Psi$ which is non-minimally coupled to other matter fields and also
to the KK scalar $\Phi$. The action for the second scalar $\Psi$ is
given by\footnote{The original action of \cite{ML} also included a
self-interaction term of the $U(\Psi)$. This term is however later in
\cite{ML} discarded as small. Thus we decide to neglect it in our
considerations right from the beginning for conciseness.}
\begin{equation}
S_\Psi = \int d^4x\, \sqrt{-g} \, \Phi \, \left[ \frac 12 \d_\mu \Psi \d^\mu \Psi - W(\Psi,\Phi,\ldots)
\right] \, ,
\label{scalaraction}
\end{equation}
where the term $W$ represents the non-minimal coupling of $\Psi$ to other
fields.  Mbelek and Lachi\`eze-Rey make the ansatz that the interaction term
$W$ can be expressed as a sum of interaction terms with different fields, all
of which however also depend on the KK scalar $\Phi$,
\begin{equation}
W(\Psi,\Phi,\ldots) = W(\Phi,\Psi) + W(\Phi,\Psi,A_\mu) 
+ W (\Phi,\Psi,T_\text{matter}) \, ,
\label{fullcoupling}
\end{equation}
In Eq.\ (\ref{fullcoupling}) $A_\mu$ denotes the electromagnetic vector
potential and $\Phi$ is assumed to couple to conventional matter only via a
coupling to the trace of its stress-energy tensor $T_\text{matter} \equiv
g_{\mu\nu} T^{\mu\nu}_\text{matter}$.  The interaction with the
  electromagnetic field is assumed to take the form
\begin{equation}
W (A_\mu,\Psi) = w_\text{EM}(\Phi,\Psi)\, \epsilon_0 F_{\mu\nu} F^{\mu\nu} \, ,
\label{EMcoupling}
\end{equation}
where the coupling $w_\text{EM}$ is a function of both scalar fields.  The
couplings of $\Phi$ to other fields remain unspecified but are assumed to be
negligible in the weak-field approximation. The contribution $S_\text{matter}
= \int d^4x \sqrt{-g} \Phi \C{L}_\text{matter}$ in Eq.\ (\ref{ac1}) represents the
action of other matter fields with the usual minimal coupling to the
effective metric
$\Phi^2 g_{\mu\nu}$.
 
Surprisingly, the action $S_\text{``KK''}$, Eq.\ (\ref{MLKKaction}), features
a kinetic term for the scalar which is usually absent in the effective KK
action in Jordan frame (cf. e.\,g.\ 
\cite{Ravndal,AppelquistIntro}),\footnote{Conventions are metric signature
  $(+,-,-,-)$ and positive sign of the Einstein tensor.}
\begin{equation}
S_\text{KK} = \int d^4x\, \sqrt{-g} \left[ 
- \frac{c^4}{16\pi G_0} \Phi R + \frac 14 \epsilon_0 \Phi^3 F_{\mu\nu} F^{\mu\nu}
\right] \, .
\label{JordanAction}
\end{equation}
The scalar kinetic term in Eq.\
(\ref{MLKKaction}) is neither a total divergence nor can it be
generated form Eq.~(\ref{JordanAction}) by a conformal transformation.
Moreover varying the action $S_\text{``KK''}$, Eq.\ (\ref{MLKKaction})
with respect to $\Phi$ one obtains an equation of motion
which is inconsistent with the extra-dimensional component of the
five-dimensional equations of motion of KK theory in the Jordan 
frame,
\begin{equation}
R_{55}=\frac 83 \pi G_5 g_{55} g^{MN} \delta^\mu_M \delta^\nu_N
T_{\mu\nu} \, , \label{55eom}
\end{equation}
where $G_5$ is the five-dimensional Newton's constant, the $5$ denotes
the index of the compactified extra dimension, $M,N = 0,\dots,3,5$,
and $T_{\mu\nu}$ is the stress-energy tensor of four-dimensional
matter including the scalar $\Psi$. The equation of motion for the
KK scalar $\Phi$, given in, Eq.\ (11) of \cite{ML}, has the form one
would 
obtain from the higher-dimensional component of the Ricci tensor
amended by the terms one obtains by varying the action of the second
scalar $\Psi$, Eq.\ (\ref{scalaraction}), with respect to
$\Phi$. However, this would not be the correct
equation of motion for $\Phi$ even if one would replace
$S_\text{``KK''}$, Eq.\ (\ref{MLKKaction}), by Eq.\ (\ref{JordanAction}).

Another problem with the action presented in \cite{ML} is that the
kinetic terms of the KK scalar $\Phi$ and the other scalar
$\Psi$ enter with different signs (cf.\ Eqs.\ (\ref{MLKKaction}) and
(\ref{scalaraction})).  Actually, in \cite{ML} it is concluded that a
\emph{classical} instability is associated with the KK scalar $\Phi$
which is in contrast to all previous treatments of KK
theory.\footnote{For a didactic derivation and discussion of the KK
action in various frames see e.\,g.\ \cite{Ravndal}.} In order to
locate the source of this misconception we analyse the conventions
employed in \cite{ML}. By calculating the stress-energy tensor for
both the electromagnetic and the scalar part of the action, Eq.\
(\ref{MLKKaction}), we find that the kinetic term of the KK scalar has
the correct sign relative to the action of the electromagnetic field.
Hence if the KK scalar in the action, Eq.\ (\ref{MLKKaction}) were a
ghost, as claimed in \cite{ML}, then also the energy of the
electromagnetic field in the model would be negative.

The first two terms of the action, Eq.\ (\ref{MLKKaction}), just yield the
usual KK action displayed for a metric with signature $(+,-,-,-)$ and
negative sign of the Einstein tensor\footnote{ There is no way to
deduce the sign of the Riemann tensor from Eq.\ (\ref{MLKKaction}).}
\begin{equation}
 G_{\mu\nu} = - 8\pi G T_{\mu\nu} \, ,
\end{equation}
where $G_{\mu\nu} = R_{\mu\nu} - \frac 12 g_{\mu\nu} R$ denotes the Einstein
tensor.  On the other hand, from the effective Einstein equations given in \cite{ML}, there Eq.\ 
(4), and the kinetic term of the scalar $\Psi$ we infer the
conventions: metric signature $(+,-,-,-)$ and 
\begin{equation}
G_{\mu\nu} = + 8\pi G T_{\mu\nu} \, .
\end{equation}
Consequently the action of \cite{ML} is a mixture of contributions written in
different conventions leading to the misinterpretation that KK theory is
\emph{classically} unstable.\footnote{The classical stability properties of
  KK theory are discussed in a didactic way in \cite{Blagojevic} 298
  ff. The alleged ghost in \cite{ML}
is not related to the ghost appearing in KK theory with
\emph{more than one} extra dimension in \cite{Sokolowski}. The scalar ghost
discussed in the latter article arrises similar to the conformal ghost in
ordinary Einstein theory (cf.\ e.\,g.\ \cite{Mazur}) by singling out the
conformal mode of the extra-dimensional hypersurface.} 
In order to
keep our reasoning as close as possible to the investigations of \cite{ML} but
nevertheless avoid the inconsistency in sign we change the ``KK'' part of
the action, (Eq.~\ref{MLKKaction}), 
to so-called Landau-Lifschitz timelike conventions,\footnote{The name referring
  to the book \cite{LLTL}.} i.\,e.\ metric signature
$(+,-,-,-)$ and $G_{\mu\nu} = + 8\pi G T_{\mu\nu}$. The action, Eq.\
(\ref{MLKKaction}), then becomes
\begin{equation}
S_{g\Phi F} = \int d^4x\, \sqrt{-g} \left[ 
- \frac{c^4}{16\pi G_0} \Phi R + \frac 14 \epsilon_0 \Phi^3 F_{\mu\nu} F^{\mu\nu}
+ \frac{c^4}{4\pi G_0} \frac{\d_\mu \Phi \, \d^\mu \Phi}\Phi
\right] \, ,
\label{myKKaction}
\end{equation}
and the total modified action becomes
\begin{equation}
S = S_{g\Phi F} + S_\Psi + S_\text{matter} \, ,
\label{totaction2}
\end{equation}
where $S_\Psi$ is given by Eq.\ (\ref{scalaraction}) and $S_\text{matter}$ is
the usual action of additional matter fields. 

By conducting only the above minimal modification
to render the action of \cite{ML} consistent we unfortunately have to
drop the idea
that the effective action under study can be derived by the
KK mechanism from a five-dimensional theory. 
For an effective theory derived by KK reduction we would have to 
employ the effective action, Eq.\ (\ref{JordanAction}) amended by the
$55$-component of the five-dimensional Einstein equations. In this
case the $55$-component of the Einstein-equations would take the role
of the equations of motion for the KK scalar $\Phi$,
\begin{equation}
\Box \Phi = \frac 14 \epsilon_0 \Phi^3 F_{\mu\nu} F^{\mu\nu}+
\frac 83 \pi G_5 \Phi^3 g^{\mu\nu} T_{\mu\nu} \, .
\label{KKPhieom}
\end{equation}
In Eq.\ (\ref{KKPhieom}) the coupling of the KK scalar $\Phi$ to matter goes
with the same power of $\Phi$ than that to the electromagnetic field.
Consequently, in this case the contribution to the gravielectric coupling from
the $F_{\mu\nu} F^{\mu\nu}$ term in $S_{KK}$, Eq.\ (\ref{JordanAction}), will
be of the same order of magnitude as that from the term, Eq.\ 
(\ref{EMcoupling}), and the effective theory will look very different from the
model studied by Mbelek and Lachi\`eze-Rey.  In contrast to this the suggested
sign changes in the action do only marginally affect the phenomenological
discussion of \cite{ML} and render the action consistent at least as a
four-dimensional theory.

Interestingly, an action differing from Eq.\ (\ref{MLKKaction}) only by a
numerical coefficient has been presented as the action of KK theory in Jordan
frame once before the works of Mbelek and Lachi\`eze-Rey in
\cite{Wesson}. However, this is an obvious misprint because \cite{Appelref},
which is cited as the source of the action in \cite{Wesson}, does not display
it but is only concerned with the KK action in Einstein frame.

Obviously, the sign with which the kinetic term of the scalar appears in the
KK action in Einstein frame is tied to the conventions employed. This is
necessary because the relative sign of the Einstein tensor and the
stress-energy tensor in the Einstein equations also depends on the conventions
and one should not have the stability properties of the effective action
changed by changing conventions.  It is worth understanding the causes which
in \cite{ML} led to the assumption of a \emph{classical} instability of KK
theory from this point of view. They are already rooted in Mbelek's and
Lachi\`eze-Rey's earlier article \cite{MLghost}. However, as this topic lies
somewhat out of the main scope of our considerations we defer the analysis of
\cite{MLghost} to the appendix.

In order to investigate the phenomenological viability of the
minimally improved model based on \cite{ML} we derive in the
following, the linearised equations of motion of the action, Eq.\
(\ref{totaction2}) which will enable us to study phenomenological
aspects of the weak-field regime of the theory.

\section{The linearised equations of motion \label{lin}}

Already in \cite{ML} the equations of motion for the Lagrangian, Eq.\ 
(\ref{ac1}), have been given and used to study the weak field regime of the
model, in that case in particular the field configuration around the Earth.
However, in \cite{ML} no consistent linearisation of the action was
conducted but instead the full equations of motion where simplified by
determining small terms from plausibility arguments and dropping these from
the field equations.  In order to put our analysis on a more systematic
footing we linearise each term of the action in the fields
$g_{\mu\nu},\Phi,\Psi$ to lowest non-vanishing order and afterwards derive the
equations of motion from this linearised action. This procedure seems
especially desirable because below we will encounter an error in the equations
of motion as displayed in \cite{ML} (and also in \cite{MLfine,MLnew}).

Following \cite{ML} we assume that the scalar $\Phi$ far
from any matter sources takes the value $\Phi = 1$ and thus can be expanded as
\begin{equation}
\Phi = 1 + \varphi + O(\varphi^2)\, .
\label{philin}
\end{equation}
The field $\Psi$ is assumed to asymptotically take the value $v$, thus
having the expansion
\begin{equation}
\Psi = v + \psi + O(\psi^2) \, .
\end{equation}
Around the Minkowski background the metric has the expansion
\begin{equation}
g_{\mu\nu} = \eta_{\mu\nu} 
 + \frac{\sqrt{32\pi G_0}}{c^2} \left(1- \frac \varphi2\right) h_{\mu\nu} +
O (h^2,\varphi^2) \, , 
\end{equation}
where we have included a factor $\sqrt{32 \pi G_0/\Phi}/c^2$ in order
to ensure the symmetry of the linearised theory under time reversal ---
or in more practical terms the symmetry of the linearised gravitational
interaction between two masses under the interchange of the masses
--- and we have made use of Eq.\ (\ref{philin}).  Expanding each
term to lowest non-vanishing order in $\varphi$ \emph{and}
$h_{\mu\nu}$ we find for the KK part of the action
\begin{multline}  
S_{g\Phi F\text{,lin}} = \int d^4 x
\left[ \frac{c^4}{16\pi G_0} \C{L}_\text{FP} 
+  \frac{\sqrt{8\pi G_0}}{c^2}\left(1- \frac \varphi2\right) h_{\mu\nu} T^{\mu\nu}_\text{EM} \right. \\
\left. + (1+ 3\varphi) \frac{\epsilon_0}4 F_{\mu\nu}F^{\mu\nu}
+ \frac{c^4}{4\pi G_0} \d_\mu \varphi \d^\mu \varphi
\right] \, .\label{KKlin}
\end{multline}
where $\C{L}_\text{FP}$ 
denotes the usual Fierz-Pauli Lagrangian of the linearised 
gravitational field,
\begin{equation}
\C{L}_\text{FP} = \frac{32 \pi G_0}{c^4}
\left[
\frac 12 h_{\mu\nu,\sigma} h^{\mu\nu,\sigma}
- h_{\mu\nu,\sigma}h^{\mu\sigma,\nu}
+ h_{\mu\sigma}\vphantom{h}^{,\sigma}h^{,\mu}
- \frac 12 h_{,\nu}h^{,\nu}
\right] \, ,
\end{equation}
with $h = \eta^{\mu\nu}h_{\mu\nu}$, and $T^{\mu\nu}_\text{EM}$ in Eq.\ 
(\ref{KKlin}) denotes the stress-energy tensor of the electromagnetic field.

To determine the linearised action of the scalar fields $\Phi$, $\Psi$ we first note
the expansion of $W_\text{EM}(\Phi,\Psi,A_\mu)$,
\begin{align}
W_\text{EM}(\Psi,\Phi) &= W_\text{EM}(v,1) 
+\left. \frac{\d W_\text{EM}}{\d\psi}\right|_{\genfrac{}{}{0pt}{}{\psi=0}{\varphi=0}} \psi
+\left. \frac{\d W_\text{EM}}{\d\varphi}\right|_{\genfrac{}{}{0pt}{}{\psi=0}{\varphi=0}} \varphi
+ O(\phi^2,\varphi^2,\psi\varphi) \nn \\
& = \epsilon_0 F_{\mu\nu} F^{\mu\nu} \left( 
       w_\text{EM}(v,1) 
+ \left. \frac{\d w_\text{EM}}{\d\psi}\right|_{\genfrac{}{}{0pt}{}{\psi=0}{\varphi=0}} \psi
+\left. \frac{\d w_\text{EM}}{\d\varphi}\right|_{\genfrac{}{}{0pt}{}{\psi=0}{\varphi=0}} \varphi
   \right) \nn \\ &\hspace{6cm} + O(\phi^2,\varphi^2,\psi\varphi)\, .
\label{coupling}
\end{align}
Using Eq.\ (\ref{coupling}) and assuming in accord with \cite{ML} that
the couplings $W(\Phi,\Psi)$ and $W(T,\Phi,\Psi)$ in Eq.\
(\ref{fullcoupling}) are negligible we find as the action for the
linearised scalar $\psi$
\begin{multline}
S_{\psi\text{,lin}} = \int d^4 x  
\Bigg[ \frac 12 \d_\mu \psi \d^\mu \psi \\
- \epsilon_0 F_{\mu\nu} F^{\mu\nu} \left(  w_\text{EM}(v,1) + 
\left. \frac{\d w_\text{EM}}{\d\psi}\right|_{\genfrac{}{}{0pt}{}{\psi=0}{\varphi=0}} \psi
+\left. \frac{\d w_\text{EM}}{\d\varphi}\right|_{\genfrac{}{}{0pt}{}{\psi=0}{\varphi=0}} \varphi
   \right)
\Bigg] \, . \label{psilin}
\end{multline}

The total linearised action is composed from Eqs.\ (\ref{KKlin}),
(\ref{psilin}) and the linearised action of matter fields
\begin{equation}
S_\text{lin} = S_{g\Phi F\text{,lin}} + S_{\psi\text{,lin}}+
S_{\text{matter,lin}} \, . \label{linaction}
\end{equation}
Due to the contribution, Eq.\ (\ref{coupling}), the electromagnetic field
couples non-minimally to $\psi$ and the theory remains non-metric even
in the linearised approximation in $\Phi$ and $\Psi$. As a particular
phenomenological consequence we expect violations of the WEP. Since
the WEP is tested to high accuracies a first question to the model
of \cite{ML} should be if it respects current experimental
bounds on violations of the WEP.  We thus calculate the modifications
to be expected from the action, Eq.\ (\ref{linaction}), compared to
four-dimensional Einstein theory for typical E\"otv\"os experiments
with test bodies of different materials.

In order to proceed in this direction we first obtain the equations
of motion for $\varphi$ and $\psi$ by varying the linearised action, Eq.\
(\ref{linaction}),
\begin{align}
&\Box \varphi + \frac{2 \pi G_0\, \epsilon_0}{c^4} F_{\mu\nu}F^{\mu\nu} \,
\left. \frac{\d w_\text{EM}}{\d\varphi}\right|_{\genfrac{}{}{0pt}{}{\psi=0}{\varphi=0}}
= 0 \, ,
\label{phieom}
\\
&\Box \psi + \epsilon_0 F_{\mu\nu}F^{\mu\nu} \,
\left. \frac{\d w_\text{EM}}{\d\psi}\right|_{\genfrac{}{}{0pt}{}{\psi=0}{\varphi=0}} =0 \, ,
\label{psieom}
\end{align}
where
\begin{equation}
F_{\mu\nu}F^{\mu\nu} = 2 \, \left(\frac{{\bf B}^2}{c^2} - {\bf E}^2\right)
\end{equation}
with ${\bf B}$ and ${\bf E}$ being the 3-vectors of the electric and magnetic
field, respectively and we have neglected a term of $O(h_{\mu\nu})$ in Eq.\ 
(\ref{phieom}). 

The equations of motion for the metric perturbation are also obtained from
varying the action, Eq.\ (\ref{linaction}). In harmonic gauge, 
$h^{\mu\nu}\vphantom{h}_{,\mu} - \frac 12 h_{,\nu} = 0$, they simplify to
\begin{multline}
- \left(
\eta_{\mu\lambda}\eta_{\nu\rho} +
\eta_{\mu\rho}\eta_{\nu\lambda} - \eta_{\mu\nu}\eta_{\lambda\rho}
\right) \Box h^{\lambda\rho}(x_\alpha) 
= \\
\frac{\sqrt{8\pi G_0}}{c^2}
\left[1-\frac{\varphi(x_\alpha)}2\right] 
[ T_\text{EM}^{\mu\nu}(x_\alpha) +  T_\text{matter}^{\mu\nu}(x_\alpha)] \, .
\label{heom}
\end{multline}
In Eq.\ (\ref{heom}) the stress energy tensors of the scalars do not appear
because they would enter the gravitational equations only at third order in the
linearised fields $\varphi,\psi,h_{\mu\nu}$.

Taking the static limit  of the 
Eqs.\ (\ref{phieom}) and (\ref{psieom}) we find
\begin{align}
\triangle \varphi &= \frac{4 \pi G_0}{c^4} 
\left. \frac{\d w_\text{EM}}{\d\varphi}\right|_{\genfrac{}{}{0pt}{}{\psi=0}{\varphi=0}}
\, \epsilon_0\left( \frac{{\bf B}^2}{c^2} - {\bf E}^2 \right) \, ,
\label{staticphi} \\
\triangle \psi &=
\left. \frac{\d w_\text{EM}}{\d\psi}\right|_{\genfrac{}{}{0pt}{}{\psi=0}{\varphi=0}}
\,  \epsilon_0 \left( \frac{{\bf B}^2}{c^2} - {\bf E}^2 \right) \, . 
\label{staticpsi}
\end{align}
Comparing our corrected equation of motion for the scalar $\varphi$ with the
dipole fit to the magnetic field of the Earth given in \cite{ML} we obtain the
numerical value
\begin{align}
\frac{2 \pi G_0}{c^4} 
\left. \frac{\d w_\text{EM}}{\d\varphi}\right|_{\genfrac{}{}{0pt}{}{\psi=0}{\varphi=0}}
&= - (5.44 \pm 0.66) \, 10^{-6} \, \frac{\text{fm}}{\text{TeV}}\nonumber \\
&= - (3.40 \pm 0.41) \, 10^{-8} \, \text{m}/\text{J} \, .
\end{align}
In the following section we will use this value of the coupling to
analyse the strength of the violations of the WEP in E\"otv\"os
experiments in the model of \cite{ML}.

Finally, we have for the gravitational field of a static source
\begin{align}
\triangle h^{00} (\vec{x})
&= \frac{\sqrt{8\pi G_0}}{c^2}
\left(1-\frac{\varphi(\vec{x})}2\right) 
\left[T_\text{EM}^{00}(\vec{x}) +  T_\text{matter}^{00}(\vec{x})\right] \,,
\nn \\
h^{ij}(\vec{x}) &= \delta^{ij} h^{00}(\vec{x}) \, ,
\quad i,j=1,2,3\, ,
\quad \quad \quad h^{\mu 0}(\vec{x})=0 \, ,
\label{statich}
\end{align}
where we have assumed that the pressure of the source can be neglected,
$T^\mu\vphantom{T}_\mu = T^{00}$.

\section{Violations of the WEP \label{WEP}}

As already emphasised above the theory based on the action of
Eq.~(\ref{totaction2}) or (\ref{linaction}) cannot be cast into a form in
which a combination of the metric and the $\Phi$ scalar couples universally to
matter.  As a consequence of this non-metricity the theory will feature
violations of the WEP. In order to be a viable theory these violations must
obey current experimental bounds.  The violations of the WEP will manifest
themselves when electromagnetic fields are present. Since a strong electric
field is present in the nucleus of every atom already the data from
conventional E\"otv\"os experiments provide a first challenge to the theory.

In order to derive an estimate on the strength of violation of the WEP
in an E\"otv\"os experiment we employ a simple model of the atomic
nucleus in which the mass density and the charge density of a specific
element are assumed to be constant throughout the nucleus.
Neglecting the effect of the electron sheath, the electric field of
the nucleus is then given by
\begin{equation}
E (\vec{r}) = \frac{eZ\, \vec{e}_r}{4\pi \epsilon_0} \left\{ \begin{array}{ll}
\displaystyle \frac 1{r^2} \, ,& r > R \\
\displaystyle \frac r{R^3} \, ,& r \leq R  \, .
\end{array} \right.
\label{eofnucl}
\end{equation}
Here $\vec{e}_r$ is the radially pointing unit vector, $e$ is the charge of the
proton, $Z$ denotes the atomic number of the nucleus and $R$ denotes the
radius of the core which is related to the nucleon number, $A$, as
\begin{equation}
R = r_0 A^{1/3} \, , 
\label{RofA}
\end{equation}
with $r_0 = 1.3 \pm 0.1 \times 10^{-13}\, {\rm m}$ being an empirical
value (see e.\,g.\ \cite{MayerKuckuk}).
In order to find the field configuration for the linearised scalar 
$\varphi$ around a nucleus we
note that the solution of the Poisson equation
\begin{equation}
\triangle \varphi(\vec{r}) = f(r)
\end{equation}
for a spherically symmetric potential $f(\vec{r})\equiv f(r)$, $r=|\vec{r}|$, 
extending to a radius $r=X$, is given by 
\begin{equation}
\varphi(\vec{r}) = \int_0^X d^3 r' \frac{f(r')}{|\vec{r} - \vec{r'}|}
\, . \label{poissonint}
\end{equation}
This can be brought into the form
\begin{equation}
\varphi (\vec{r}) =
\frac{4\pi}r \, 
\left\{ \begin{array}{ll}
\displaystyle \int_0^X dr' \left[ {r'}^2 f(r') \right] \, ,
& \text{for }r > X \\
\displaystyle \int_0^rdr' \left[ {r'}^2 f(r') \right] 
+ \int_r^X dr' \left[ rr' f(r') \right] \, ,
& \text{for }r \leq X  \, ,
\end{array}
\right.
\label{poissonsol}
\end{equation}
In our case, inserting Eq.\ (\ref{eofnucl}) into Eq.\  (\ref{staticphi}), the potential $f(r)$ under consideration is
\begin{equation}
f(r) =
\frac{G_0 \, e^2 Z^2}{4 \pi \epsilon_0 c^4} \, 
\left. \frac{\d w_\text{EM}}{\d \varphi} \right|_{\genfrac{}{}{0pt}{}{\psi=0}{\varphi=0}}
\left\{ \begin{array}{ll}
\theta(-X) \, 1/r^4\, , &r > R\\ 
r^2/R^6\, , &r \leq R \, ,
\end{array} \right. \label{fofr}
\end{equation}
where we have introduced a step function $\theta(X)$ as a primitive
approximation for the shielding of the electric field outside of the nucleus
by the electron sheath. Despite its assumed temperature dependence (cf.\ 
\cite{ML}) the coupling parameter $\d w_\text{EM} / \d \varphi |_{\psi =0,
  \varphi = 0}$ is a good approximation of the one obtained from the fit to the
Earths magnetic field, because the temperature to be associated with a nucleus
in its ground state is $T=0$ and nearly all baryonic
matter on Earth is in its ground state.
For the time being we will not specify the
radius $X$ where the field drops to zero but instead keep it as a free
parameter.

In the expression for the gravitational field of the nucleus (see below)
the scalar field $\varphi(\vec{r})$ will be folded with
the mass density of the nucleus $\rho(\vec{r})$ which is zero outside the
core. Hence, only the case
\begin{equation}
r \leq R < X
\end{equation}
is relevant for our considerations,
i.\,e.\ we are only interested in the value of the scalar within the radius of
the nucleus and the electric shielding by the electrons is assumed to take
effect only outside of the core. Thus, the integral to solve becomes
\begin{equation}
\varphi(\vec{r}) 
= \frac{G_0 \, e^2 Z^2}{\epsilon_0 c^4 r} 
\left. \frac{\d w_\text{EM}}{\d \phi} \right|_0
\left(
  \int_0^r d r' \frac{{r'}^4}{R^6}
  + \int_r^X d r' \frac{r {r'}^3}{R^6}
  + \int_R^X d r' \frac{r}{{r'}^3}
\right) \, .
\label{phiint}
\end{equation}
Evaluating Eq.\ (\ref{phiint}) one finds
\begin{align}
\varphi (\vec{r}) =
\frac{G_0 \, e^2 Z^2}{\epsilon_0 c^4} 
\left. \frac{\d w_\text{EM}}{\d \varphi} \right|_{\genfrac{}{}{0pt}{}{\psi=0}{\varphi=0}}
\left( \frac 34 \frac1{R^2} -\frac1{20}\frac{r^4}{R^6}-\frac 1{2X^2}
\right) \, .
\end{align}
In the following we can neglect the contribution from the last term in the sum
because the shielding of the electric field by the electron sheath takes
effect only far outside of the nucleus, i.\,e.\ we have $X \gg R$ for any
element. Dropping the term $\sim X^{-2}$ we arrive at the final expression for
the field of the linearised scalar $\varphi$ around a nucleus of atomic number
$Z$ and nucleon number $A$,\footnote{There is no reason to expect the
  behaviour $\varphi \to \infty$ for $r \to 0$ as claimed in \cite{Mcomment}
  because $r=0$ is not an asymptotic point of the domain of $\varphi$. Hence
  nothing drives the solution of the differential equation Eq.\ 
  (\ref{staticphi}) towards the solution of the homogeneous equation
  $\triangle \varphi = 0$.}
\begin{equation}
\varphi (r) = \varphi_0
\left(\frac1{20}\frac{Z^2 r^4}{r_0^4 A^2} - \frac 24 \frac{Z^2}{A^{2/3}}
\right) \, ,
\label{phiofnucleus}
\end{equation}
where we have used Eq.\ (\ref{RofA}) and
defined a constant
\begin{equation}
\varphi_0 \equiv \frac{G_0 \, e^2}{\epsilon_0 c^4 r_0^2} 
\left. \frac{\d w_\text{EM}}{\d \varphi} \right|_{\genfrac{}{}{0pt}{}{\psi=0}{\varphi=0}} \, .
\end{equation}

As a next step we have to find the metric perturbation generated by 
a nucleus in the
presence of the field Eq.\ (\ref{phiofnucleus}). Using Eq.\ (\ref{poissonint})
on Eq.\ (\ref{statich}) we find for the
$00$-component of the metric-perturbation of a single nucleus
\begin{equation}
h^{00}(\vec{r}) = - \sqrt{8\pi G_0} \int_0^R d^3r' 
\left[ \left(1-\frac{\varphi(\vec{r'})}2\right)
\frac{\rho (\vec{r'})}{|\vec{r}-\vec{r'}|} \right]\, ,
\label{nonrelh1}
\end{equation}
where we have used that the $00$-component of the
stress-energy tensor for the nucleus at rest is
given by $T_{00} = \rho(\vec{r'})c^2$.
As a reasonable approximation we can assume that the density of the nucleus is
constant $\rho (\vec{r'}) = \rho_0$. 
Hence, at a distance $r>R$ from the centre of the nucleus its effective
gravitational mass is given by
\begin{equation}
M_G = r \,\rho_0 \int_0^R d^3r' \left[ \left(1-\frac{\varphi(\vec{r'})}2\right)
\frac 1{|\vec{r}-\vec{r'}|} \right]\, .
\label{effmass}
\end{equation}
Note that this expression is not related to the gravitational self-energy of
the nucleus.\footnote{This was erroneously implied in \cite{Mcomment}.}
Instead we have simply absorbed the influence of $\varphi$ on the interaction
between gravitation and matter into the definition of an effective mass, Eq.
(\ref{effmass}). The back-reaction of $\varphi$, as well as of the metric
field $h_{\mu\nu}$, on the mass of the nucleus is not considered as it is
of higher order in the perturbation expansion.  Hence a calculation of the
self-energy contribution to the E\"otv\"os experiment is beyond the scope of
the linearised approach.\footnote{The self-energy contribution to a violation
  of the WEP is expected to be of the order of $10^{-25}$ and thus far beyond
  present experimental limits (cf.\ \cite{Nordtvedt}).}

Since we are going to check for the experimental limits of an E\"otv\"os experiment
using a torsion balance with different materials we are only interested in the
metric perturbation outside of the material, i.\,e.\ only
the case $r > R \ge r'$.

Violations of the equivalence principle are usually not formulated in
terms of the metric perturbation generated by the test bodies but in
terms of a dimensionless quotient of the effective gravitational and
the inertial mass of the matter, $M_G$ and $M_I$,\footnote{cf.\
\cite{Will} p.\ 24 ff.\ for an introduction to tests of the WEP and
\cite{Willlife} for an overview on recent experimental results.}
\begin{equation}
a = M_G/M_I \, .
\end{equation}
From our expression for the effective mass, Eq.\ (\ref{effmass}),
we easily deduce this ratio for our case
\begin{equation}
M_G = \frac{ \displaystyle{ \int_0^R d^3r' 
\frac{1-\varphi(\vec{r'})/2}{|\vec{r}-\vec{r'}|}}}{4\pi R^3/3r} \, ,
\end{equation}
where the constant density of the nucleus $\rho_0$ has cancelled out.
Evaluating the integral with the help of Eq.\ (\ref{poissonsol}) we find
\begin{equation}
a = 1 - \frac {51}{70} \frac {\varphi_0 Z^2}{A^{2/3}} \, .
\label{anucl}
\end{equation}
The parameter $a$ determined in this way is only valid for the nucleus without
an electron sheath. Outside of the nucleus the shielding of the electric
field by the electrons takes effect. Hence also the scalar $\varphi$ which
has the electric field as its source will be decaying (cf.\ Eq.\ 
(\ref{fofr})). Both fields will not change sign outside the nucleus.
Therefore the change to the gravitational mass of the electron sheath will have
the same sign as that of the nucleus. In order not to consider the complicated
structure of the electric field outside the nucleus we resort to the limit of
no violations of the WEP for the electron sheath, yielding
the smallest violation of the WEP for the whole atom and thus yielding a lower
bound on $|a-1|$ for the atom. Under this assumption $a$ for the atom is given
by
\begin{equation}
a = \frac{M_{G,\text{nucl}}+M_e}{M_{I,\text{nucl}}+M_e}\, ,
\label{aatom}
\end{equation}
where $M_{G,\text{nucl}}$ and $M_{I,\text{nucl}}$ are the gravitational and
inertial mass of the nucleus, respectively, and $M_e$ denotes the mass of the
electrons of the sheath. Expanding this expression for $M_e$ and using Eq.\
(\ref{anucl}) we find to first order in $M_e$
\begin{equation}
a = 1  - \frac {51}{70} \frac {\varphi_0 Z^2}{A^{2/3}} 
\left[ 1 - \frac{Zm_e}{m_p}\right]
\, ,
\end{equation}
where $m_e$ and $m_p$ are the mass of the electron and the proton,
respectively. An E\"otv\"os experiment will test the parameter $a$ not for a
single atom but for a macroscopic body of a specific material. If the reaction
of the torsion balance to the gravitational field of a distant object is
studied the test bodies can be considered as point masses and the parameter
$a$ of the macroscopic body can be obtained by averaging over the parameters
$a$ of the single atoms. Hence for a macroscopic body consisting of one
chemical element $a$ is found by taking for $A$ the average nucleon number of
the element.

In a torsion-balance experiment the gravitational force exerted by two 
different
materials is compared. The quantity extracted from the experiments is 
\begin{equation}
\eta \equiv \frac{2|a_1 - a_2|}{|a_1 + a_2|} \, .
\end{equation}

To make a comparison between our theoretical estimate and the two classic
E\"otv\"os experiments in Moscow \cite{Braginsky} and Princeton \cite{Roll} as
well as the more recent one in Seattle \cite{Su} we calculate the coefficients
$a$ for the chemical elements used for the test bodies in these experiments.
The results are given in Table \ref{aforexp}.

\begin{table}
\begin{center}
\begin{tabular}{c@{~}r@{~}r@{}l@{~}r@{.}l@{~}r}
Element& $Z$ &\multicolumn{2}{c}{$A$}&\multicolumn{2}{c}{$(A/Z^{(1/3)})^2$}&
\multicolumn{1}{c}{$a-1$}\\
\hline
Be& 4 & 9&& 3&70 & $-1.3\pm0.2 \times 10^{-9}$\\
Al& 13 & 27&& 18&78 & $-6.3\pm 1.0 \times 10^{-9}$ \\
Cu& 29 & 63&.6& 52&77 & $-1.8\pm0.2 \times 10^{-8}$\\
Pt& 78 & 195&.1& 180&86 & $-5.9\pm 1.0 \times 10^{-8}$ \\
Au& 79 & 197& &~~~~~184&34 & $-6.0\pm 1.0 \times 10^{-8}$ \\
\hline
\end{tabular}
\caption{Determination of the nucleonic contribution to the parameter $a$ for
  beryllium, aluminium, cooper, platinum and gold. The average values for the
  nucleon numbers were calculated from the data in \cite{Isotopes}.
\label{aforexp}}
\end{center}
\end{table}

From the values for $a$ we obtain the parameter $\eta$ for the material
combinations of the  Moscow \cite{Braginsky} (Al and Au), Princeton
\cite{Roll} (Al and Pt) and the two Seattle experiments
\cite{Su} (Be and Al,Cu), respectively,
\begin{align}
&\eta_{\text{Al--Au}} = 5.7 \pm 0.9 \times 10^{-8}\, ,
\quad
\eta_{\text{Al--Pt}} = 5.5 \pm 0.9 \times 10^{-8}\, , \nn \\
&\eta_{\text{Be--Al}} = 5.2 \pm 0.8 \times 10^{-9}\, ,
\quad
\eta_{\text{Be--Cu}} = 1.7 \pm 0.3 \times 10^{-8}\, .
\label{etatheo}
\end{align}
The errors for these results are dominated by the uncertainty in the radius
of the nuclei i.\,e. by uncertainty of the parameter $r_0$. All other
uncertainties are several orders of magnitude below this contribution. Thus,
the value of the parameter $\eta$ is rather sensitive to the nuclear model
employed. Nevertheless, the above values give some reasonable impression of the
size of $\eta$ to be measured in E\"otv\"os experiments.
 
Noting that the Moscow experiment has constrained the parameter
$\eta_{\text{Al--Au}}$ to $10^{-12}$ one finds that the violations to the WEP
occurring in the model given by the action of Eq.\ (\ref{totaction2}) are four
orders of magnitude above the experimental limits for the Moscow experiment.
With this huge discrepancy we can safely conclude that even the most intricate
model of the nucleus will non alter our basic conclusion about the
non-viability of the theory described by the action, Eq.\ (\ref{totaction2}).
This result also holds true for the original action of \cite{ML}, including
the contribution Eq.\ (\ref{MLKKaction}) instead of Eq.\ (\ref{myKKaction}),
because the sign change of the kinetic term of the scalar field only amounts
to a sign change in the parameters $a-1$, leaving unaffected the numerator of
$\eta$ and thus only slightly modifies the above results for $\eta$.

The model is also in conflict by several orders of magnitude, with the other
two experiments which had obtained $\eta_{\text{Al--Pt}} \lesssim 10^{-11}$
and $\eta_{\text{Be--Al}} \lesssim 10^{-12}$, $\eta_{\text{Be--Cu}} \lesssim
10^{-12}$.
Thus, we can rule out the strong gravielectric coupling mediated by two
scalar fields as suggested by Mbelek and Lachi\`eze-Rey
 in \cite{ML} because it generates violations to
the WEP far beyond the experimental bounds.

\section{Conclusions \label{concl}}

We have studied some aspects of the model proposed in \cite{ML}. 
The model is claimed to consists
of five dimensional KK action which is reduced to four dimensions applying the
usual cylinder condition. To this effective action which already includes
electromagnetism and a KK scalar, the action of the other known matter fields
and a further scalar are added. The second scalar couples to the KK scalar in
a non-minimal way (in the five-dimensional theory). This coupling
seems rather unnatural from the point of view of unification because it
singles out the KK scalar compared to the other components of the
five-dimensional metric. More importantly the coupling between both
scalars and the electromagnetic field induces a non-metricity in the action
(which cannot be removed by any conformal redefinition of the metric).  The
model proposed in \cite{ML} is thus not comprised in the set of
biscalar-tensor theories investigated by \cite{Damour} which is restricted to
metric theories.

Unfortunately, we found that the action of the model as given in
\cite{ML} cannot be deduced from five-dimensional gravity amended
with additional fields by a KK compactification 
because it features a kinetic term for the KK scalar which is
incompatible with the higher-dimensional Einstein equations of KK
theory. 
It does not seem possible to change the model in a simple way to make
it derivable from a five-dimensional theory by KK reduction
without loosing the essential phenomenology of a gravielectric
coupling which maintains astrophysical constraints. 

In addition to this we realized that the structure of the action as
given in \cite{ML} is inconsistent in the sense that it would
yield a negative energy for the electromagnetic field.  In order to
study the phenomenology of a model which is self-consistent we thus have
investigated a modified action
with different signs for both the four-dimensional
the electromagnetic and ``KK''-scalar terms,
Eq.\ (\ref{myKKaction}), which eliminates all ghosts in the theory.

The non-metricity of the model directly leads to one test of its
phenomenology: The violations to WEP induced by the non-metricity have to stay
below current observational bounds. Since the model of \cite{ML} is required
to exhibit a strong coupling between the gravitational and electromagnetic
fields mediated by a scalar the model violates the experimental bounds
and is thus ruled out. We have demonstrated this fact by estimating the static
field of the scalar around a nucleus and then deriving from this result the
effective gravitational mass of the nucleus. We found that the violations of
the WEP generated in the model are ruled out from E\"otv\"os experiments by
several orders of magnitude. The strongest bound comes from the classic Moscow
experiment by Braginsky and Panov \cite{Braginsky} for which the violations of
the WEP predicted by the model of \cite{ML} are found to be too large by four
orders of magnitude. As a consequence of its violation of the WEP the model
will also feature violations of the Einstein Equivalence principle: As the
equations of the model are derivable from an action this directly follows from
the proof of \cite{LightmanLee}.

In conclusion it is not possible to explain the discrepancy of the various
measurement of Newton's constant by assuming a sufficiently strong
gravielectric coupling. This conclusion seems to be generic because the method
to test the gravielectric coupling against constraints from torsion-balance
experiments does not rely on the specific realisation of the coupling. It is
always possible to determine the effective gravitational mass of test bodies
from the linearised static configuration of the field, which mediates the
gravielectric coupling, around the body. Since the baryonic matter of the test
bodies in torsion-balance experiments is in the same state as the typical
constituents of the Earth the test body will be exposed to a gravielectric
coupling with the same coupling constant as experienced by the Earth as a
total.  A coupling of the gravitational field to the Earth's magnetic field of
a strength of $10^{-3}$ will by the covariance of the electromagnetic field
equations be accompanied by a strong coupling of the gravitational field to
the electric field of atomic nuclei and is thus bound to generate
experimentally non-viable violations of the WEP even if the model manages to
escape astrophysical constraints.

Of course, the results found above do neither rule out scalar fields in
general nor exclude that they may play an important role in the
cosmological evolution.  However, in the same sense as the binary-pulsar data
yielded strong constraints on the Brans-Dicke parameter describing the
coupling between gravitation and a scalar field (cf.\ e.\,g.\ \cite{Willlife})
our considerations put tight constraints on the possible coupling strength
between a scalar field and the electromagnetic field. In particular they rule
out that a scalar-mediated coupling of the electromagnetic and gravitational
field can explain the discrepant measurements of Newton's constant.

\paragraph*{Acknowledgement}
The author is grateful to J.-P. Mbelek for helpful discussions and
explanations.

\begin{appendix}

\section*{Appendix: No scalar ghost in KK theory}

In their paper \cite{MLghost} Mbelek and Lachi\`eze-Rey claim that the
effective four-di\-men\-sio\-nal action of KK theory contains a scalar
ghost in its spectrum. This claim is justified by an explicit
decomposition of the five-dimensional Ricci scalar and thus the five
dimensional Einstein-Hilbert action. In order to understand how their
misconception arrises we take a quick tour through the appendix of
\cite{MLghost} with special emphasis on the conventions employed.

The appendix of \cite{MLghost} uses a metric of signature $(+,-,-,-)$ and
defines the (five-dimensional) Ricci scalar as
\begin{equation}
R = g^{AB} \left(
\d_N \G^N_{AB} - \d_B \G^N_{AN}
+ \G^N_{AB} \G^M_{NM} - \G^M_{AN} \G^N_{BM}
\right) \, 
\end{equation}
(Latin indices run from 0 to 4, Greek indices run from 0 to 3).
From the definition of the Ricci scalar we see that their Einstein equations
take the form
\begin{equation}
G_{MN} = + 8\pi G_5 T_{MN}
\end{equation}
and the gravitational and matter parts of their action thus must have opposite
signs,  
\begin{equation}
S= - \frac1{16\pi G_5} \int d^5x \sqrt{-g} R 
+ \int d^5x \sqrt{-g} \C{L}_\text{matter} \, .
\label{apaction} 
\end{equation}
Choosing a specific gauge and KK decomposing the Ricci scalar Mbelek and
Lachi\`eze-Rey find as the contribution of the KK scalar to the
action\footnote{ Actually, in the decomposition we encounter another error in
  the calculations of \cite{MLghost}: despite of choosing a gauge in which the
  components $g_{4\mu}$, are set to zero, they still have derivatives of these
  metric components present in their final decomposed Ricci scalar. In
  contrast to the claim in the footnote 5 of \cite{MLghost} there is of course
  no way of having a function vanishing everywhere with a non-vanishing
  derivative.}
\begin{equation}
\vphantom{V}^{(5)}R = \vphantom{V}^{(4)}R \ldots 
+ \frac 12 \frac {\d_\mu\Phi \d^\mu \Phi}\Phi \, .
\end{equation}
Since the kinetic term for $\Phi$ has the inverse sign compared to that of a
normal scalar field for the above metric signature it is concluded in
\cite{MLghost} that the KK scalar is a ghost signalling a classical instability
of the KK action.

However, due to the sign difference between the gravitational part and the
matter part of the action, cf.\ Eq.\ (\ref{apaction}), the KK scalar will
appear in the action with the same sign as an ordinary scalar field. Thus, the
KK scalar does not have a ghost nature and does not signify any instability of
five-dimensional gravity. In conclusion, the scalar ghost occurred to Mbelek and
Lachi\`eze-Rey only because they did not keep track of their conventions and
it is not present in the physical theory. It is needless to say that none of
the articles \cite{Maheshwari,Collins,AppelquistIntro}, referred to in \cite{Mcomment} as supporting the presence of an
instability, mention any stability problem of the KK action.\footnote{These
  articles employ various conventions but are always consistent with
  stability in the sense that the kinetic term of a fourdimensional scalar
  added by hand to the KK action has the same sign as the KK scalar arising
  from dimensional reduction.}

\end{appendix}

\end{document}